\documentstyle[12pt,aps,epsf,amsfonts,amssymb]{revtex}
\def\mycap#1{ 
\parbox[h]{\textwidth}{\vskip 0.4cm
\footnotesize \baselineskip 0.mm  #1 } }

\begin{document}
\author{$^{1}${\bf Miodrag} {\bf L.} {\bf Kuli\'{c} and }$^{2}${\bf Igor M. Kuli\'{c}%
}}
\address{$^{1}$Physikalisches Institut, Theorie III, \\
Universit\"{a}t Bayreuth, 95440 Bayreuth, Germany\\
$^{2}$Institut f\"{u}r Theoretische Physik 1, Universit\"{a}t Stuttgart,\\
70563 Stuttgart, Germany}
\title{{\bf Josephson Effect in Magnetic Superconductors\\ 
with Spiral Magnetic Order%
}}
\date{01 January 2000 - revised version 25 Mai 2000}
\maketitle

\begin{abstract}
It is shown that in magnetic superconductors with spiral magnetic order the
Josephson current has an additional contribution which depends: (i) on the
relative orientation (magnetic phase) $\theta =\theta _{L}-\theta _{R}$ of
magnetizations on the left ($L$) and right ($R$) banks of the contact, (ii)
on the junction helicity $\chi =\chi _{L}\chi _{R}$, (with 
spiral helicity $\chi _{L(R)}=\pm 1$) , i.e. $J=[J_{c}-J_{\chi }\cos
\theta ]\sin \varphi $ with $\varphi =\varphi _{L}-\varphi _{R}$. The
ratio $R_{\chi }\equiv J_{\chi }/J_{c}$ is calculated as a function of the
superconducting order parameter $\Delta $, the exchange field energy $h$ and
the wave vector $Q$ of the spiral magnetic structure. The $\pi $-Josephson
contact can be realized in such a system in some region of parameters. Some
possible consequences of this new phase relation is also analyzed.
\end{abstract}

\newpage

\ {\it Introduction} - The physics of magnetic superconductors (MSC) is
interesting due to competition of magnetic order and singlet
superconductivity (SC) in bulk materials. The problem of their coexistence
was first studied theoretically by V. L. Ginzburg \cite{Ginzburg} in 1956,
while the experimental progress begun after the discovery of ternary rare
earth (RE) compounds (RE)Rh$_{4}$B$_{4}$ and (RE)Mo$_{6}$X$_{8}$ (X=S,Se) 
\cite{Maple} with regular distribution of localized RE magnetic ions. 
It turned out that in many of these systems SC, with the critical
temperature $T_{c}$, coexists rather easily with antiferromagnetic (AF)
order, with the critical temperature $T_{AF}$, where usually one has $%
T_{AF}<T_{c}$\cite{Maple}. Due to their antagonistic characters singlet SC
and ferromagnetic (FM) order can not coexist in bulk samples with realistic
physical parameters, but under certain conditions the FM order, in the
presence of SC, is transformed into a spiral or domain-like structure -
depending on the anisotropy of the system \cite{BuKuRu}, \cite
{BuBuKuPaAdvances}. These two orderings coexist in a limited temperature
interval $T_{c2}<T<T_{m}$ (reentrant behavior in ErRh$_{4}$B$_{4}$ and HoMo$%
_{6}$S$_{8}$) or even down to $T=0$ $K$ (in HoMo$_{6}$Se$_{8}$). Note, in
ErRh$_{4}$B$_{4}$ one has $T_{c}=8.7$ $K$, $T_{m}\approx 0.8$ $K$, $%
T_{c2}\approx 0.7$ $K$, while for HoMo$_{6}$S$_{8}$ one has $T_{c}=1.8$ $K$, 
$T_{m}\approx 0.74$ $K$, $T_{c2}\approx 0.7$ $K$ - see Refs.\cite{Maple}, 
\cite{BuBuKuPaAdvances}, \cite{BuKuRu}. In most new quaternary compounds
(RE)Ni$_{2}$B$_{2}$C the AF order and SC coexist \cite{Chang}, while in HoNi$%
_{2}$B$_{2}$C, an additional oscillatory magnetic order exists competing
strongly with SC leading to reentrant behavior \cite{DetBraun}. 
Recently, the coexistence of SC and nuclear magnetic order was observed in
AuIn$_{2}$ \cite{Pobell}, where $T_{c}=0,207$ $K$ and $T_{m}=35$ $\mu K$,
which was explained in terms of spiral or domain-like magnetic structure 
\cite{KuBuBu}. There are evidences for the coexistence of the AF (or FM)
order and SC ($T_{AF}=137$ $K$, $T_{c}<45$ $K$) 
in layered perovskite superconductor RuSr$_{2}$%
GdCu$_{2}$O$_{8}$ \cite{Braun}, where AF and SC orders are spatially
separated in Ru-O and Cu-O planes, respectively.

In all of the above MSC systems the exchange(EX) interaction between
localized magnetic moments (LM's) and SC is much larger (the EX model) than
the electromagnetic (EM) one. The latter is due to the orbital effect of the
magnetic induction ${\bf B}\approx 4\pi {\bf M}$ on SC electrons.

In the following we study the Josephson effect in MSC with spiral magnetic
order in the framework of the EX model. It is shown below that the Josephson
current depends additionally on the magnetic phase $\theta =\theta
_{L}-\theta _{R}$ of the magnetic order \ parameters, as well as on the
spiral helicities $\chi _{L(R)}=\pm 1$ on the left and right surface.

The microscopic theory of MSC \cite{BuBuKuPaAdvances} takes into account the
interaction between LM's and conduction electrons which goes via: $a)$ the
direct EX interaction and $b)$ the EM interaction where the dipolar magnetic
field ${\bf B}_{m}({\bf r})=4\pi {\bf M}({\bf r})$ created by LM's acts on
the orbital motion of electrons. The Hamiltonian of the system has the form

\[
\hat{H}=\hat{H}_{0}+\hat{H}_{BCS}+\int d^{3}r\{\hat{\psi}^{\dagger }({\bf r})%
\hat{V}_{ex}({\bf r})\hat{\psi}({\bf r})+ 
\]

\begin{equation}
+\frac{[{\rm curl}{\bf A}({\bf r})]^{2}}{8\pi }\}+\sum_{i}[-{\bf B}({\bf r}%
_{i})g\mu _{B}{\bf \hat{J}}_{i}].  \label{hamilton}
\end{equation}
$\hat{H}_{0}\equiv \hat{H}_{0}({\bf \hat{p}}-\frac{e}{c}{\bf A})$ describes the
motion of quasiparticles in the magnetic field ${\bf B}({\bf r})={\rm curl}%
{\bf A}({\bf r})$, which is due to LM's and screening superconducting
current. $\hat{H}_{BCS}\equiv \hat{H}_{BCS}(\Delta ({\bf r}))$ is the BCS
pairing Hamiltonian with the SC order parameter $\Delta ({\bf r})$, $\hat{V}%
_{ex}(r)={\bf \hat{h}(r)\sigma }$ is the EX potential where ${\bf \hat{h}(r)}%
=\sum_{i}J_{ex}({\bf r}-{\bf r}_{i})(g-1){\bf \hat{J}}_{i}$. Here, $J_{ex}(%
{\bf r})$ is the exchange integral between electronic spins ${\bf \sigma }$
and LM's and ${\bf \hat{J}}_{i}$ is the total angular momentum operator of
the LM at the $i$-th site and ${\bf \sigma }=(\sigma _{1},\sigma _{2},\sigma
_{3})$ are Pauli spin matrices. In case of magnetic anisotropy the
crystal-field term $\hat{H}_{CF}({\bf \hat{J}}_{i})$ should be added to $%
\hat{H}$.

In the following a clean $s-wave$ MSC is considered with spiral magnetic
order with the wave vector ${\bf Q}$ along the $z$-axis, ${\bf Q}=Q_{z}{\bf 
\hat{z}}=\pm Q{\bf \hat{z}}$, and the spiral helicity $\chi =Q_{z}/Q=\pm 1$.
The LM's are assumed to lie in the $xy$-plane due to the easy-plane magnetic
anisotropy and the mean-field EX potential $\hat{V}({\bf r})\equiv <\hat{V}%
_{ex}({\bf r})>$ reads 
\begin{equation}
\hat{V}({\bf r}){\bf =}\left( 
\begin{array}{cc}
0 & he^{-i(\chi Qz{\bf +\theta })} \\ 
he^{i(\chi Qz{\bf +\theta })} & 0
\end{array}
\right) .  \label{hsigma}
\end{equation}
Here, $h=n_{m}(g-1)J_{ex}(0)<\mid {\bf \hat{J}}\mid >$ and $J_{ex}(0)$ is
the $q=0$ Fourier component of $J_{ex}({\bf r}-{\bf r}_{i})$, $n_{m}$ is the
concentration of regularly distributed LM's - see \cite{BuBuKuPaAdvances}.
For further purposes we define $h_{\theta }=he^{i{\bf \theta }}$ \ where the
magnetic phase $\theta $ characterizes the orientation of the EX field at
the surface $z=0$. The Josephson current between two MSC, both with spiral
magnetic order, depends on anomalous Green's functions $F_{\sigma _{1}\sigma
_{2}}^{\dagger }(x_{1},x_{2})=<\hat{T}\hat{\psi}_{\sigma _{1}}^{\dagger
}(x_{1})\hat{\psi}\dagger _{\sigma _{2}}(x_{2})>$ $\ $with $x=({\bf r},\tau
) $ and $\sigma =\uparrow ,\downarrow $, which are calculated in \cite
{BuKuRu} 
\begin{equation}
F_{\downarrow \uparrow }^{\dagger }({\bf k},{\bf k}^{\prime };\omega
_{n})=\delta ({\bf k}-{\bf k}^{\prime })\Delta \frac{\omega _{n}^{2}+\Delta
^{2}-h^{2}+(\varepsilon _{{\bf p}}+\rho _{{\bf k}})^{2}}{[\omega
_{n}^{2}+E_{1{\bf k}}^{2}][\omega _{n}^{2}+E_{2{\bf k}}^{2}]}.  \label{Fsing}
\end{equation}
\begin{equation}
F_{\uparrow \uparrow }^{\dagger }({\bf k}+{\bf Q},{\bf k}^{\prime };\omega
_{n})=-\delta ({\bf k}-{\bf k}^{\prime })2\Delta h_{\theta }^{\ast }\frac{%
[i\omega _{n}-\rho _{{\bf k}}]}{[\omega _{n}^{2}+E_{1{\bf k}}^{2}][\omega
_{n}^{2}+E_{2{\bf k}}^{2}]}.  \label{Ftrip}
\end{equation}
The excitation spectrum $E_{1,2{\bf k}}$ is given by $E_{1,2{\bf k}%
}^{2}=\varepsilon _{{\bf k}}^{2}+\rho _{{\bf k}}^{2}+h^{2}+\mid \Delta \mid
^{2}\mp 2\sqrt{\varepsilon _{{\bf k}}^{2}(\rho _{{\bf k}}^{2}+h^{2})+\mid
\Delta \mid ^{2}h^{2}}$, where $\varepsilon _{{\bf k}}=[\xi ({\bf k}+{\bf Q}%
/2)+\xi ({\bf k}-{\bf Q}/2)]/2$ and $\rho _{{\bf k}}=[\xi ({\bf k}+{\bf Q}%
/2)-\xi ({\bf k}-{\bf Q}/2)]/2$, i.e. $\varepsilon _{{\bf k}}=\xi ({\bf k})$
and $\rho _{{\bf k}}={\bf v}_{F}\cdot {\bf Q}$. The pair function $%
F_{\downarrow \downarrow }^{\dagger }$ is obtained from $F_{\uparrow
\uparrow }^{\dagger }$ by replacing $h_{\theta }^{\ast }\rightarrow
h_{\theta }$, i.e. $\theta \rightarrow -\theta $ and ${\bf Q\rightarrow -Q}$
. In the following we assume that $Q\ll k_{F}$. Note that in the MSC 
system with
spiral magnetic order triplet pairs with amplitudes $F_{\uparrow
\uparrow }^{\dagger }$ and $F_{\downarrow \downarrow }^{\dagger }$ (which
depend on $h_{\theta }$) live short time inspite of the absence
of triplet pairing. Below is demonstrated that
the finite life-time (of the order $\hbar /\Delta $) of the triplet pairs
leads to nontrivial effects in the Josephson current.

The above theory \cite{BuBuKuPaAdvances}, \cite{BuKuRu} predicts that if the
EX interaction dominates then SC modify the FM
order into the spiral with the wave vector $Q$, ${\bf \xi }_{0}^{-1}\ll Q\ll
k_{F}$, and with zero average EX field $<{\bf h}>_{{\bf \xi }_{0}}
\approx 0$ (${\bf %
\xi }_{0}$ is the coherence length). The two ordering coexist in a narrow
temperature interval in ErRh$_{4}$B$_{4}$, HoMo$_{6}$S$_{8}$, or up to $T=0$ 
$K$ in HoMo$_{6}$Se$_{8}$. In AuIn$_{2}$ SC coexists with the nuclear
(spiral) magnetic order up to $T=0$ $K$ \cite{Pobell}, \cite{KuBuBu}.

{\it Josephson current} - We consider tunneling contact (MSC-I-MSC) between
two (left-$L$ and right-$R$) MSC both with spiral magnetic ordering with the
wave vector ${\bf Q}_{L,R}$, the SC order parameter $\Delta _{L,R}=\mid
\Delta _{L,R}\mid e^{i\varphi _{L,R}}$ and the EX field $h_{\theta
_{L(R)}}=h_{L(R)}e^{i\theta _{L(R})}$, respectively - see $Fig.1$. For
simplicity it is assumed that: (i) $\mid \Delta _{L}\mid =\mid \Delta
_{R}\mid =\mid \Delta \mid $, $h_{L}=h_{R}=h$ while $\varphi =\varphi
_{L}-\varphi _{R}\neq 0$ and $\theta =\theta _{L}-\theta _{R}\neq 0$; (ii) $%
\mid {\bf Q}_{L}\mid =\mid {\bf Q}_{R}\mid =Q$ where ${\bf Q}_{L,R}=\chi
_{L,R}Q{\bf \hat{z}}$ are orthogonal to the tunneling barrier, where
helicities of MSC are $\chi _{L(R)}=\pm 1$ for ${\bf Q}_{L,R}$ along (+) or
opposite(-) to the z-axis. In that case one has $\rho _{{\bf k,L(R)}}={\bf v}%
_{F}\cdot {\bf Q}_{L(R)}=\chi _{L(R)}\rho =\chi _{L(R)}\rho _{0}\cos \beta $%
, where $\beta \ $is the angle between ${\bf v}_{F}$ and $z$-axis.
\newline
\epsfysize=3.1in 
\hspace*{0.5cm} 
\vspace*{0cm} 
\epsffile{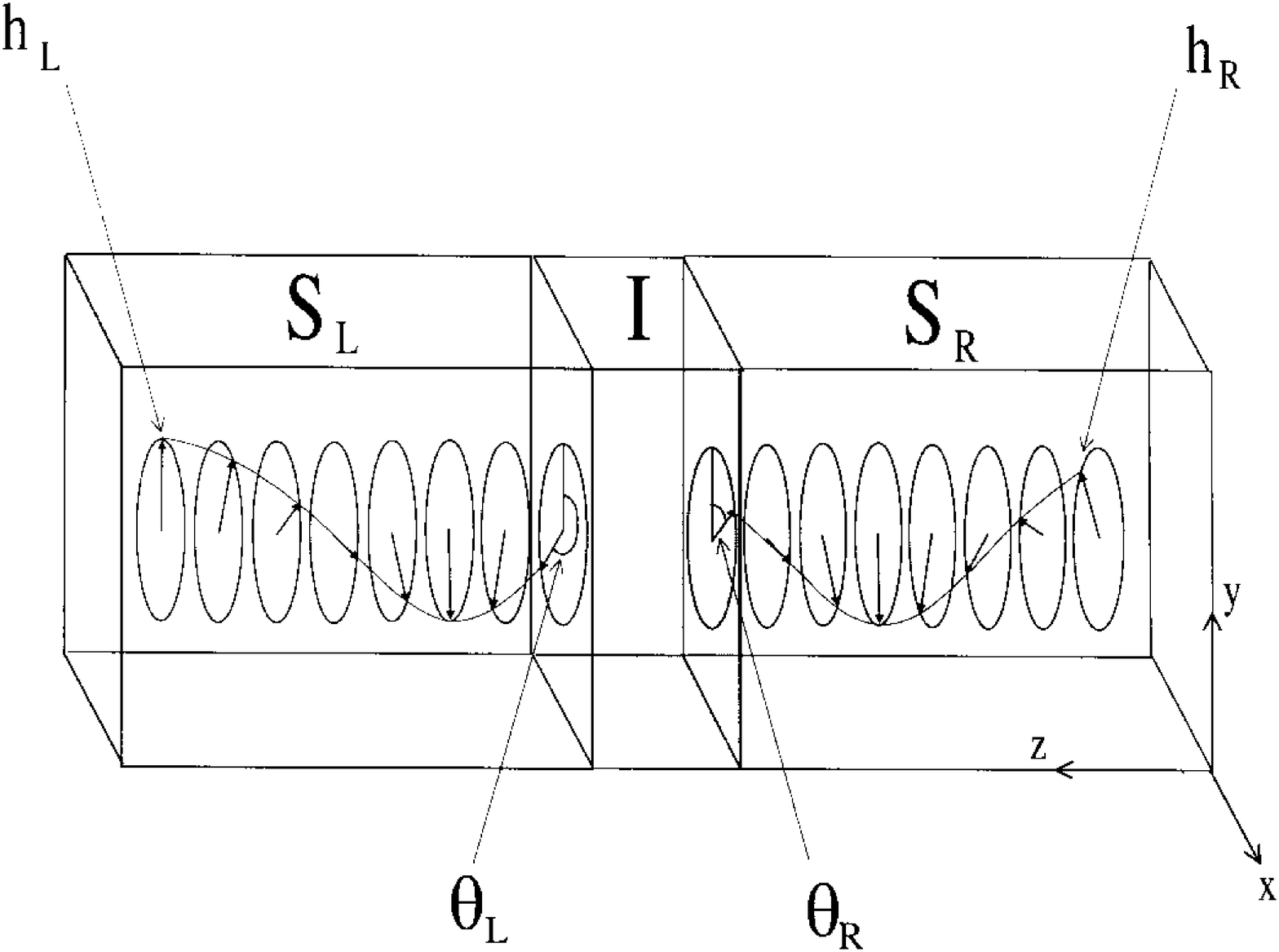} 
\newline
\mycap{{\bf FIGURE~1.} The Josephson contact $I$ 
between two magnetic superconductors $S_{L}$ and $S_{R}$
with spiral magnetic orders. The corresponding 
exchange fields
$h_{L,R}$ at the surface make angles $\theta_{L,R}$ with the $y$-axis. The
wave vectors $\bold {Q}_{L,R}$ of the spirals are along the $z$-axis}
\\

The tunneling of a left-side electron with momentum and spin (${\bf k}_{L}$, 
$\sigma $) into a right-side one (${\bf k}_{R}$, $\sigma $) is described by
the standard tunneling Hamiltonian \cite{Josep}. For simplicity it is
assumed, as usual, that the tunneling amplitude $T_{{\bf k}_{L},{\bf k}_{R}}$
is weakly energy and momentum dependent, i.e. $T_{{\bf k}_{L},{\bf k}%
_{R}}\approx T_{0}\Theta (k_{L,z}k_{R,z})$. The Heaviside function $\Theta $
takes into account that before tunneling the left electron moves toward the
barrier, while after it moves as the right electron away from the barrier,
and vice versa. The assumed $T_{{\bf k}_{L},{\bf k}_{R}}$ is more suitable
for diffusive barrier (incoherent tunneling). The extension of the theory to
the momentum dependence $T_{{\bf k}_{L},{\bf k}_{R}}$ is straightforward.

The standard theory of the Josephson effect \cite{Josep} applied to MSC
systems gives the Josephson current $J(\varphi ,\theta )$ and the 
energy of the junction $E_{J}(\varphi ,\theta )$ 
\begin{equation}
J(\varphi ,\theta )=[J_{c}-J_{\chi }\cos \theta ]\sin \varphi ,  \label{jot}
\end{equation}
\begin{equation}
E_{J}(\varphi ,\theta )=-\frac{\Phi _{0}J_{c}}{2\pi c}[1-R_{\chi }\cos
\theta ]\cos \varphi +const,  \label{ejot}
\end{equation}
where, $\varphi =\varphi _{L}-\varphi _{R}$, $\theta =\theta _{L}-\theta _{R}
$, $\Phi _{0}$ is the flux quantum and $R_{\chi }=J_{c}/J_{\chi }$. The
first term in the bracket ($\sim J_{c}$) is the standard one due to the
singlet pair tunneling $\sum_{{\bf k}_{L},{\bf k}_{R},\omega _{n}}\mid T_{%
{\bf k}_{L},{\bf k}_{R}}\mid ^{2}F_{\uparrow \downarrow ,L}^{\dagger }({\bf k%
}_{L},\omega _{n})F_{\uparrow \downarrow ,R}({\bf k}_{R},-\omega _{n})$. In
the calculation the summation over ${\bf k}$ is replaced by the integration
over $\xi $ and $\rho (\equiv \rho _{0}\cos \beta =\rho_{0}y)$, i.e. 
\begin{equation}
\sum_{{\bf k}}(...)=\frac{N(0)}{2\rho _{0}}\int^{\rho _{0}}d\rho
\int_{-\infty }^{\infty }d\xi (...)  \label{sum}
\end{equation}
$N(0)$ is the density of states on the Fermi level.${\bf \ }$After the
integration over $\xi $ one obtains $J_{c}$ (in the following $\mid \Delta
\mid \equiv \Delta $) 
\[
J_{c}=4e\pi ^{2}N^{2}(0)\mid T_{0}\mid ^{2}\Delta ^{2}T\sum_{n=1}^{\infty
}[\int_{0}^{1}I(\omega _{n},y)dy]^{2} 
\]
\begin{equation}
I(\omega _{n},y)=\frac{a_{n}+\sqrt{a_{n}^{2}-4\Delta ^{2}h^{2}}-2h^{2}}{%
\sqrt{a_{n}^{2}-4\Delta ^{2}h^{2}}\sqrt{a_{n}-2\rho _{0}^{2}y^{2}-2h^{2}+%
\sqrt{a_{n}^{2}-4\Delta ^{2}h^{2}}}}.  \label{standard}
\end{equation}
$a_{n}$ $=\omega _{n}^{2}+\rho _{0}^{2}y^{2}+h^{2}+\Delta ^{2}$, $\rho
_{0}=Qv_{F}$, $v_{F}$ is the Fermi velocity and $T$ is the temperature.

The second term in $Eq.(\ref{jot})$ ($\sim J_{\chi }$) depends on the
relative magnetic phase $\theta =\theta _{L}-\theta _{R}$ of the EX fields
at the barrier surfaces, and on the Junction helicity $\chi (\equiv \chi
_{L}\chi _{R})=\pm 1$. Both are due to the tunneling of Cooper pairs being
short time in the triplet state. $J_{\chi }$ is due to the term $\sum_{{\bf k%
}_{L},{\bf k}_{R},\omega _{n}}\mid T_{{\bf k}_{L},{\bf k}_{R}}\mid
^{2}[F_{\uparrow \uparrow ,L}^{\dagger }({\bf k}_{L},\omega
_{n})[F_{\uparrow \uparrow ,R}^{\dagger }({\bf k}_{R},-\omega _{n})]^{\ast
}+F_{\downarrow \downarrow ,L}^{\dagger }({\bf k}_{L},\omega
_{n})[F_{\downarrow \downarrow ,R}^{\dagger }({\bf k}_{R},-\omega
_{n})]^{\ast }]$. After the $\xi $-integration $J_{\chi }$ reads 
\[
J_{\chi }=16e\pi ^{2}N^{2}(0)t_{0}^{2}\Delta ^{2}h^{2}T\sum_{n=0}^{\infty
}\{[\omega _{n}\int_{0}^{1}K(\omega _{n},y)dy]^{2}+ 
\]
\[
+\chi _{L}\chi _{R}[\rho _{0}\int_{0}^{1}yK(\omega _{n},y)dy]^{2}\}, 
\]
\begin{equation}
K(\omega _{n},y)=\frac{I(\omega _{n},y)}{a_{n}+\sqrt{a_{n}^{2}-4\Delta
^{2}h^{2}}-2h^{2}}.  \label{kicir}
\end{equation}

From $Eq.(\ref{kicir})$ follows that $J_{\chi }=0$ for $h=0$, while the
standard Josephson current $J_{c}$ is finite at $h=0$ reaching its maximum. $%
J_{\chi }>0$ for the total Junction helicity $\chi (\equiv \chi _{L}\chi
_{R})=1$, while for $\chi =-1$ it can be negative depending on $\Delta
,h,\rho _{0}$ - see discussion and Figs.2b, 3b below. In order to calculate $%
J_{c}$ and $J_{\chi }$ equilibrium values of $\Delta $, $h$, $Q$, which
minimize the free-energy $F(\Delta ,h,Q)$, are needed. As it is shown in 
\cite{BuBuKuPaAdvances} $F(\Delta ,h,Q)$ (and $\Delta $, $h$, $Q$) depends
on microscopic parameters $k_{F},v_{F},n_{m},J_{ex},\Delta _{0}$, lattice
structure, etc. what shall be not studied here.

In the following the ratio $R_{\chi }(m,p)=J_{\chi }/J_{c}$ 
is calculated numerically as a function of $%
m=\Delta /h$ and $p=(\rho _{0}/h)(=Qv_{F}/h)$
at the temperature $T=0.1\Delta $. $m$ and $p$ are supposed 
to be equilibrium values. The results are shown in $Fig.2$ for $p=const$
\newline
\epsfysize=2.5in 
\hspace*{-1.0cm} 
\vspace*{0cm} 
\epsffile{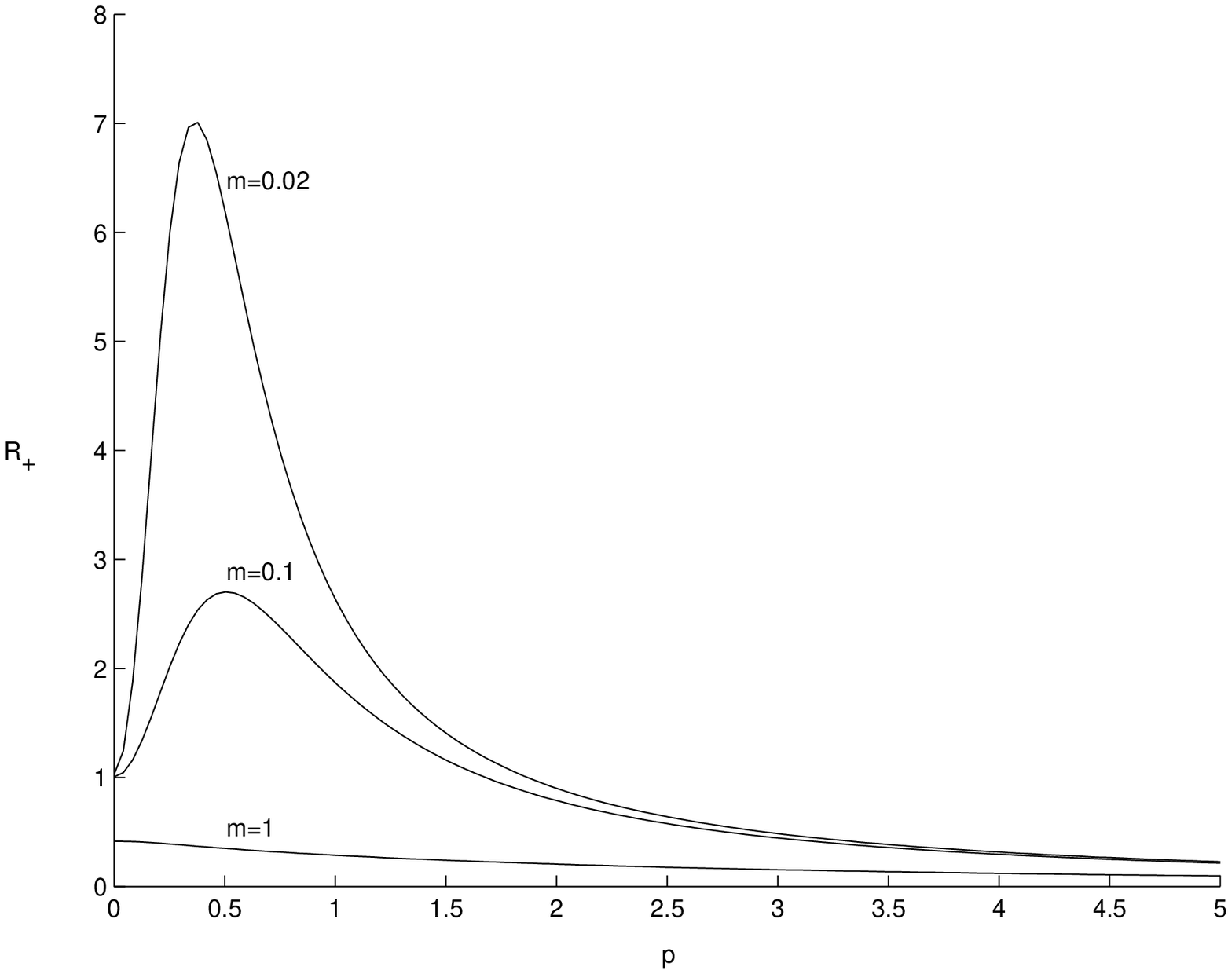}
\epsfysize=2.5in 
\hspace*{0.0cm} 
\vspace*{0cm} 
\epsffile{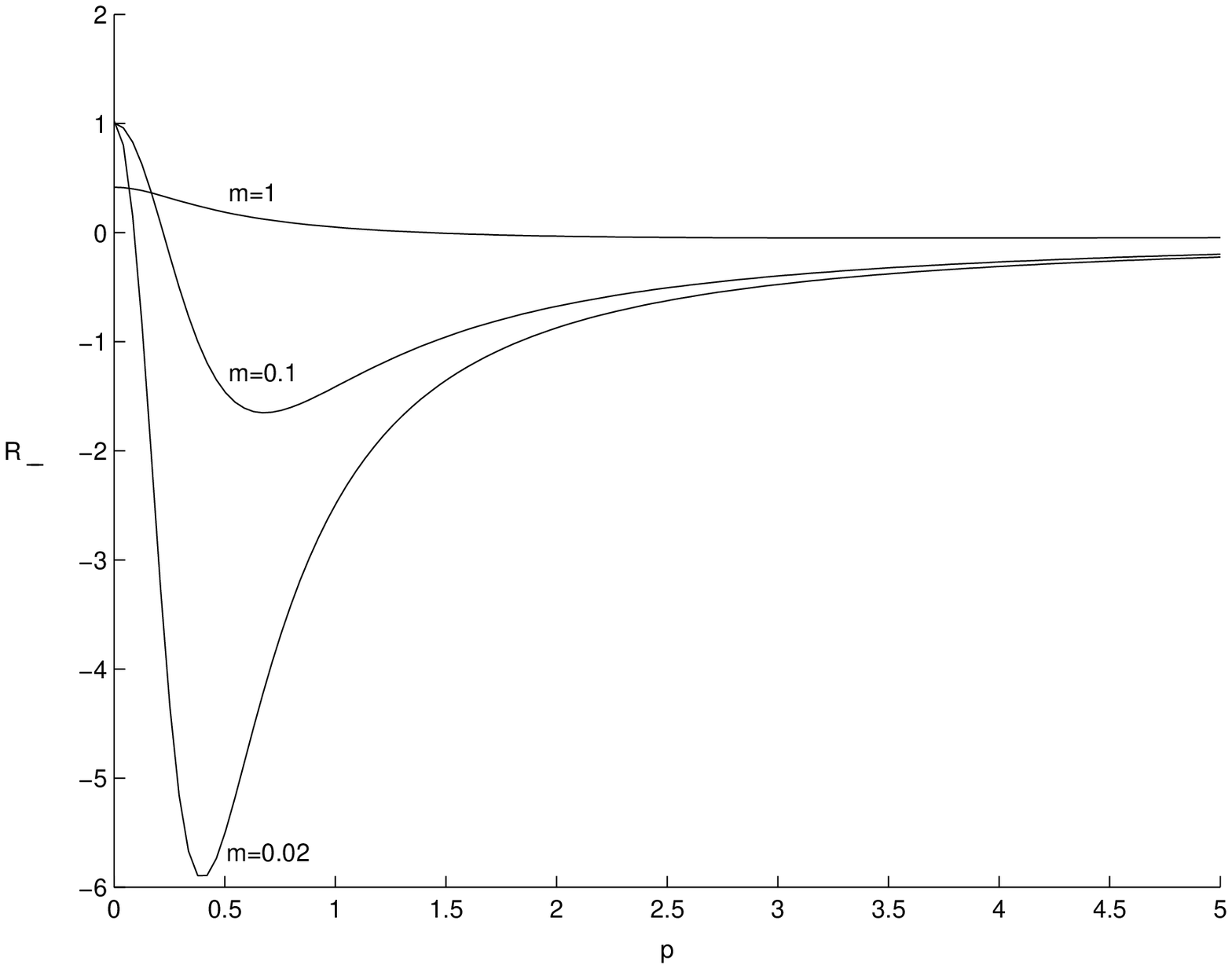}
\newline
\mycap{{\bf FIGURE~2.} The ratio $R_{\chi}(m=const,p)=J_{\chi}/J_{c}$
for various $m=1; 0.1; 0.02$. (a) ${\chi}=+1$; (b) ${\chi}=-1.$
For $\mid R_{\chi }\mid >1$ the $\pi$ - contact is realized.}\\
\\
\\
and in $Fig.3$ for $m=const$.
\\ 
\\
\epsfysize=2.5in 
\hspace*{-0.6cm} 
\vspace*{0cm} 
\epsffile{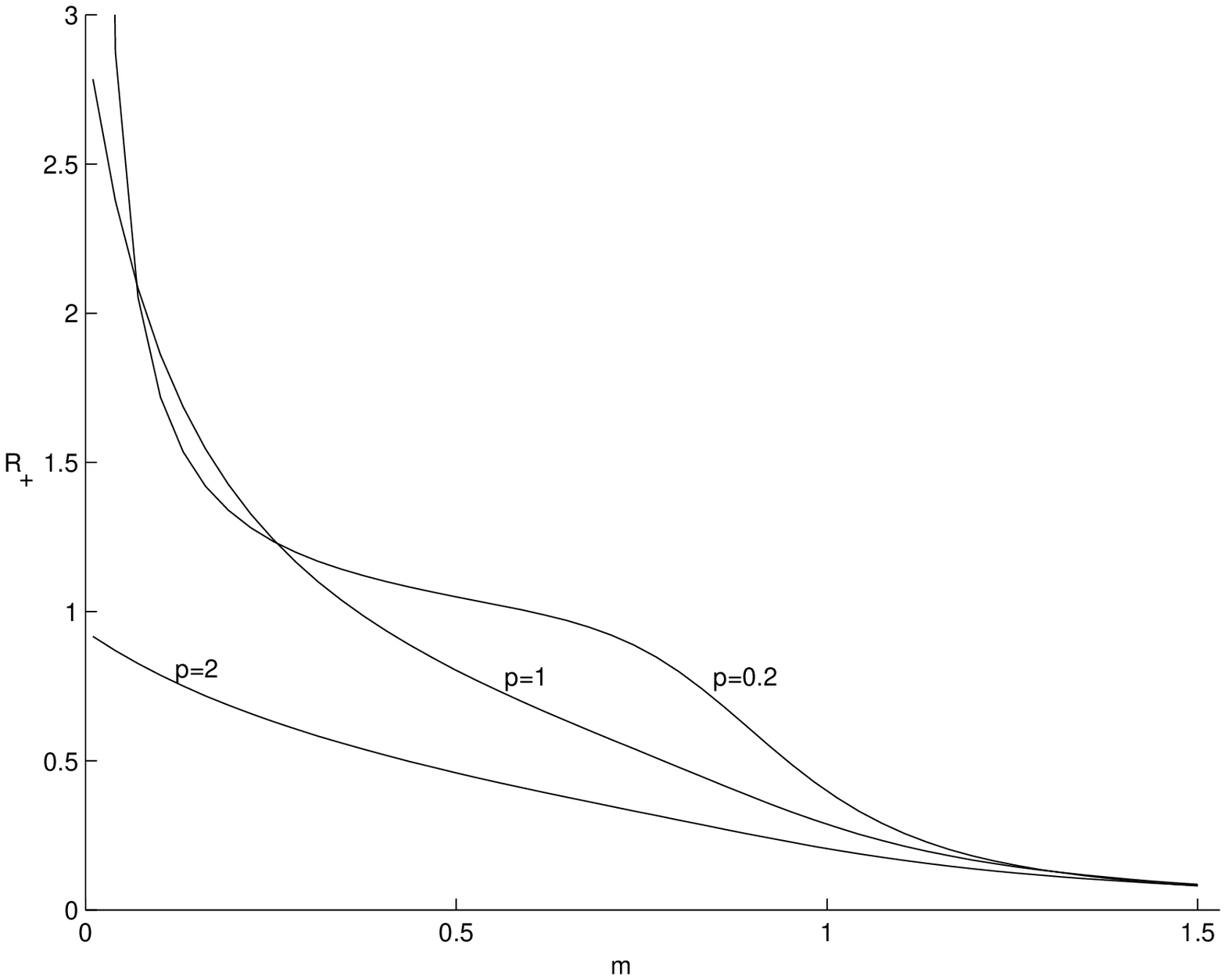}
\epsfysize=2.5in 
\hspace*{0.0cm} 
\vspace*{0cm} 
\epsffile{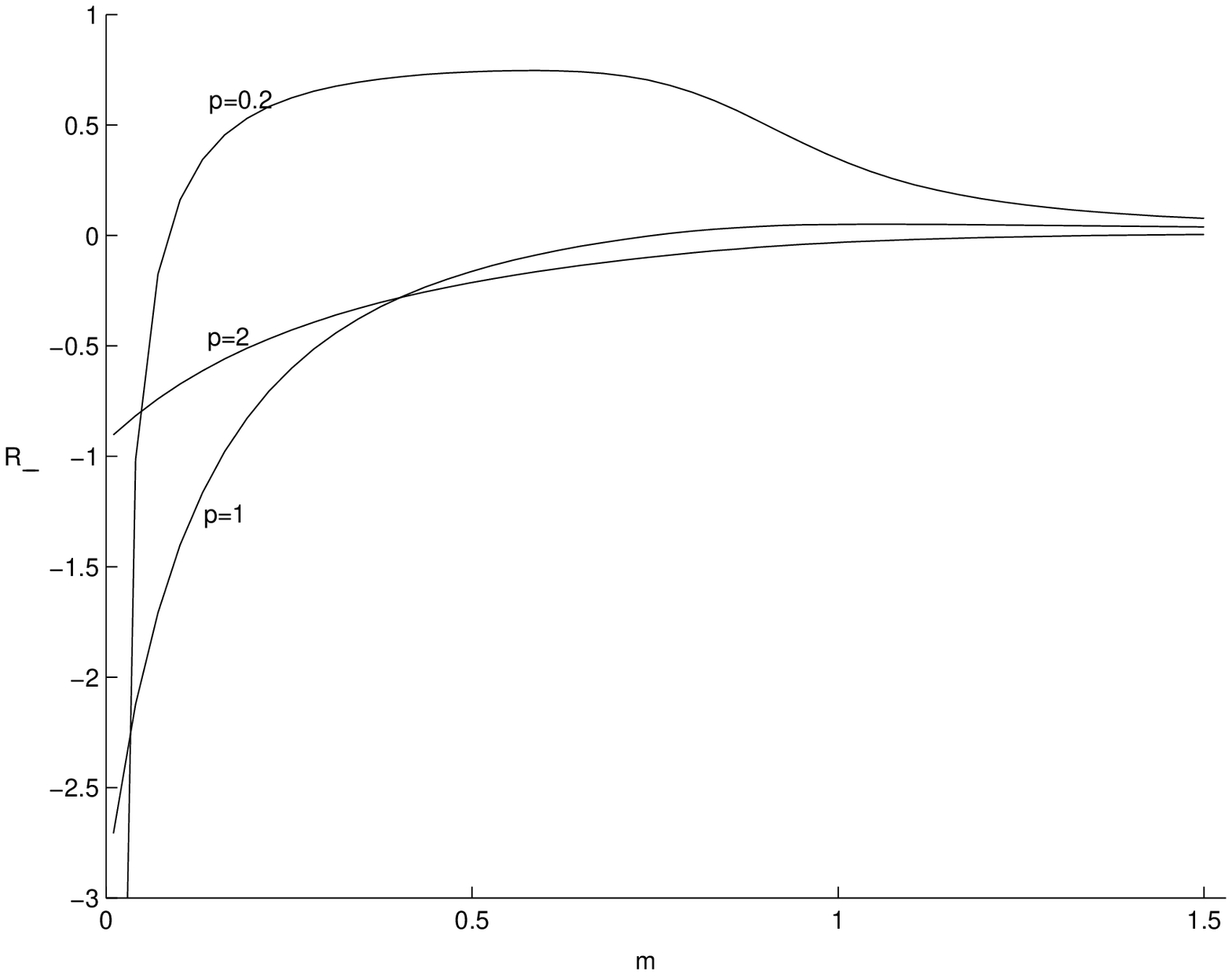} 
\newline
\mycap{{\bf FIGURE~3.} The ratio $R_{\chi}(m,p=const)=J_{\chi}/J_{c}$
for various $p=2; 1; 0.2$. (a) ${\chi}=+1$; (b) ${\chi}=-1.$
For $\mid R_{\chi }\mid >1$ the $\pi$ - contact is realized.}

{\it Discussion} 

(i) $\pi ${\it -junction}: - The above theory predicts that, besides $\mid
R_{\chi }\mid <1$, the case $\mid R_{\chi }\mid >1$ can be realized
depending on $m$ and $p$ - see $Figs.(2-3)$. From $Eqs.(\ref{jot},\ref{ejot})$
comes out that for $\mid R_{\chi }\mid >1$ the $\pi $-junction is realized,
i.e. $\min E_{J}$ is reached for $\varphi =\pi $. $Eq.(\ref{jot},
\ref{ejot})$ imply
also that for $\chi =-1$ the case $R_{\chi }<-1$ is realized in some region
of $m$ and $p$ (see Figs.(2-3)) and the $\pi $-contact is realized for those 
$\theta $ with $\cos \theta <-1/\mid R_{\chi }\mid $. For $R_{\chi }>1$ the $%
\pi $-contact is realized for $\cos \theta >1/\mid R_{\chi }\mid $. The
favorable range of parameters $m$ and $p$ for the realization of $\pi $%
-junction are seen from $Figs.(2-3)$ and it lies in the region $m<1$, $p<2$.
Note that $m(\equiv \Delta /h_{ex})$ is a measure of the strength of the SC
order parameter with respect to the magnetic order parameter (exchange
energy), while $p(\equiv Qv_{F}/h_{ex})=l_{ex}/L_{spiral}$ measures the
inverse of spiral period ($L_{spiral}=2\pi /Q$). This means that MSC systems
with spiral magnetic order which are more ferromagnetic-like are favorable
for the realization of the $\pi $-contact. On the other side the latter
property is less favorable for the coexistence of SC and spiral magnetic
order. In that respect it is worth of mentioning that in ferromagnetic SC
ErRh$_{4}$B$_{4}$, HoMo$_{6}$S$_{8}$, HoMo$_{6}$Se$_{8}$ and AuIn$_{2}$, $m$
varies from $m>1$ to $m\ll 1$ by lowering $T$ from $T_{m}$, while $p\gg 10$
is realized thus making $\mid R_{\chi }\mid $ small in these systems \cite
{BuBuKuPaAdvances}.

The proposed theory holds also for MSC with AF magnetic order (the limiting
case of the spiral order) with an easy-plane magnetic anisotropy. For
instance in some AF heavy-fermion superconductors, like URu$_{2}$Si$_{2}$
with $T_{AF}\approx 17$ $K$ and $T_{c}\approx 1.5$ $K$, one has $p\approx
1-2 $, $m\approx (0.01-0.03)$, where small value of $p$ is due to the small
Fermi velocity \cite{Sauls}. If one assumes that in this system
(anisotropic) s-wave SC is realized in absence of the AF order, then the
latter changes SC in the way described above giving rise to short living
triplet pairs, gapless superconductivity, power low behavior, etc. \cite
{KuKe}. Since in URu$_{2}$Si$_{2}$ one has $p\approx 1-2$, $m\approx
(0.01-0.03)$ and $\mid R_{\chi }\mid \gtrsim 1$ this means that it is
favorable for making $\pi $-junctions. In that respect other heavy fermions,
like UPd$_{2}$Al$_{3}$ ($T_{AF}\approx 14$ $K$, $T_{c}\approx 2$ $K$) and UNi%
$_{2}$Al$_{3}$ ($T_{AF}\approx 4.6$ $K$, $T_{c}\approx 1$ $K$), might belong
to the class described by the above theory \cite{KuKe}. UPt$_{3}$ ($%
T_{AF}\approx 5$ $K$, $T_{c}\approx 0.5$ $K$) is also a candidate for such a 
$\pi $-junction, if it can be described by the above theory, because $%
p\approx 1-2$, $m\approx (0.01-0.03)$ in it. However, various experiments in
UPt$_{3}$ are well described by the unconventional superconductivity \cite
{Sauls} - the type of order parameter is still under debate.

(ii) {\it magnetic phase }$\theta $: In the case of an easy-plane (x-y
plane) magnetic anisotropy the magnetic phase difference $\theta (\equiv
\theta _{L}-\theta _{R})$ can be tuned in the range $(\pi ,-\pi )$, for
instance, by applying magnetic field on lateral surfaces of the left and
right superconducting banks. So if the parameters ($m,p$) of the system
allows that $\mid R_{\chi }\mid >1$ then by rotating the external magnetic
field one can tune the Josephson junction from the $0$- to $\pi $-junction.
Note that contrary to the $\pi $-junction based on d-wave high-T$_{c}$%
-superconductors junctions based on MSC system can be varied from $0$- to $%
\pi $-junction by simple changing orientation of external magnetic field.
Even in the case $\mid R_{\chi }\mid <1$, when only $0$-junction can be
realized, by tuning $\theta $ one can make significant changes of the
Josephson current.

The proposed MSC-I-MSC junction opens new physical possibilities if spatial
and time variations of $\theta (x,y,t)$ are realized. It is to expect
that magnetic and electric fields, applied in the contact, will change both
phases $\theta (x,y,t)$ and $\varphi (x,y,t)$. The elaboration of these
effects in a form of coupled equations for $\varphi (x,y,t)$ (a modified
Ferrell-Prange equation) and for $\theta (x,y,t)$ is the matter of future
researches.

In conclusion, we demonstrate that in magnetic superconductors with spiral
magnetic order the Josephson current depends on new degrees of freedom: (1)
the relative magnetic phase (orientation)  $\theta =\theta _{L}-\theta _{R}$
of the exchange fields (magnetizations) on the barrier, and (2) on the
helicities of the left and right spiral, i.e. on $\chi (=\chi _{L}\chi
_{R})=\pm 1$. In some range of parameters $\Delta $, $h_{ex}$ and $Qv_{F}$
the $\pi $-junction can be realized by tuning $\theta $ in external magnetic
field. Even in the case of the $0$-junction the Josephson current can be
varied significantly by tuning $\theta $ thus giving rise to interesting
physics. Time and spatial changes of the magnetic phase $\theta $ and
the Josephson phase $\varphi $ can be of potential interest for 
small-scale applications, of course if the MSC-I-MSC junction is realizable.

{\it Acknowledgments} - M. L. K. acknowledges the support of the Deutsche
Forschungsgemeinschaft through the Forschergruppe ''Transportph\"{a}nomene
in Supraleitern und Suprafluiden''.

\end{document}